\documentclass[aps,reprint,superscriptaddress]{revtex4-1}

\usepackage{amsmath}
\usepackage[separate-uncertainty]{siunitx}
\usepackage{mhchem}

\renewcommand{\vec}[1]{\mathbf{#1}}
\newcommand{\abs}[1]{\left\vert #1 \right\vert}
\newcommand{\avg}[1]{\left< #1 \right>}

\usepackage{graphicx}
\graphicspath{{figures/}}

\setcitestyle{numbers,comma,square}

\begin{document}

\title{Multi-angle holographic characterization of individual fractal aggregates}

\author{Rafe Abdulali}
\affiliation{Packer Collegiate Institute, Brooklyn, NY 11201, USA}

\author{Lauren E. Altman}
\author{David G. Grier}
\affiliation{Department of Physics and Center for Soft Matter Research,
New York University, New York, NY 10003, USA}

\begin{abstract}
Holographic particle characterization uses quantitative analysis of
holographic microscopy data to precisely and rapidly measure the
diameter and refractive index of individual colloidal spheres in their
native media.  When this technique is applied to inhomogeneous or
aspherical particles, the measured diameter and refractive index
represent properties of an effective sphere enclosing each particle.
Effective-sphere analysis has been applied successfully to populations
of fractal aggregates, yielding an overall fractal dimension for the
population as a whole.  Here, we demonstrate that holographic
characterization also can measure the fractal dimensions of an
individual fractal cluster by probing how its effective diameter and
refractive index change as it undergoes rotational diffusion.  This
procedure probes the structure of a cluster from multiple angles and
thus constitutes a form of tomography.
Here we demonstrate and
validate this effective-sphere interpretation
of aspherical particles' holograms
through experimental studies on
aggregates of silica nanoparticles grown under a range of conditions.
\end{abstract}

\maketitle

\section{Introduction}

Many colloidal particles of natural and industrial interest are formed
by random aggregation of nanometer-scale monomeric units.  Examples
include protein aggregates in biopharmaceutical products
\cite{winters2020quantitative}, haze in beer, soot from flames,
astronomical dust particles, nanoparticle agglomerates in precision
polishing slurries \cite{cheong2017holographic} and microplastics in
the environment \cite{bianco2021identification}.  Standard optical
methods for particle characterization, such as static and dynamic
light scattering, yield population-averaged views of such particles'
structural properties \cite{lindsay1987properties,xu2001particle}.
Particle-resolved characterization measurements based on holographic
microscopy in principle could probe the structure of individual
fractal aggregates \cite{fung2019computational}, but so far have been
used to study the average properties of
populations of aggregates
\cite{wang2016fractal,wang2016protein,odete2020role}.  Here, we
demonstrate a fast and effective method to assess the structural
properties of individual colloidal fractal aggregates from sequences
of holograms recorded at multiple angles.

Each hologram of a fractal aggregate encodes information about the
particle's three-dimensional structure
\cite{wang2016protein,wang2016fractal,fung2019computational}.
Extracting that information with light-scattering theory is a
high-dimensional inverse problem \cite{perry2012real} whose numerical
convergence is both slow and uncertain.  The corresponding analysis
for a homogeneous sphere, by contrast, converges rapidly and yields
precise values for the sphere's diameter and refractive index
\cite{lee2007characterizing}.  Applying the efficient
spherical-particle analysis to a hologram of a fractal cluster yields
values for the diameter and refractive index that characterize an
effective sphere \cite{wang2016fractal,odete2020role} comprised of the
cluster itself and the fluid that fills its pores.  This approach
builds on the recent observation that effective-sphere analysis can be
used to measure the orientation of colloidal dimers in shear flows
\cite{altman2021holographic}.  For an irregularly shaped cluster,
these effective-sphere values depend on both the cluster's orientation
and also its morphology \cite{wang2016fractal,fung2019computational}.
Previous experimental studies recorded a single hologram of each
cluster in a population and combined effective-sphere results to infer
information about the particles' morphology
\cite{wang2016fractal,wang2016protein}.  This approach presumes that
all of the clusters in the population are grown under similar
conditions so that their results can be combined meaningfully.

Here, we record holographic videos of individual fractal clusters as
they undergo translational and rotational diffusion.  This protocol
yields holograms of each cluster in multiple orientations.  When
analyzed with the effective-sphere analysis, the set of measurements
provides insights into the morphology of that specific cluster.
Single particle morphological analyses then can be compared with
holographic analysis of the population as a whole performed with
single snapshots of clusters in a flowing fluid.

\begin{figure*}[ht]
  \includegraphics[width=0.75\textwidth]{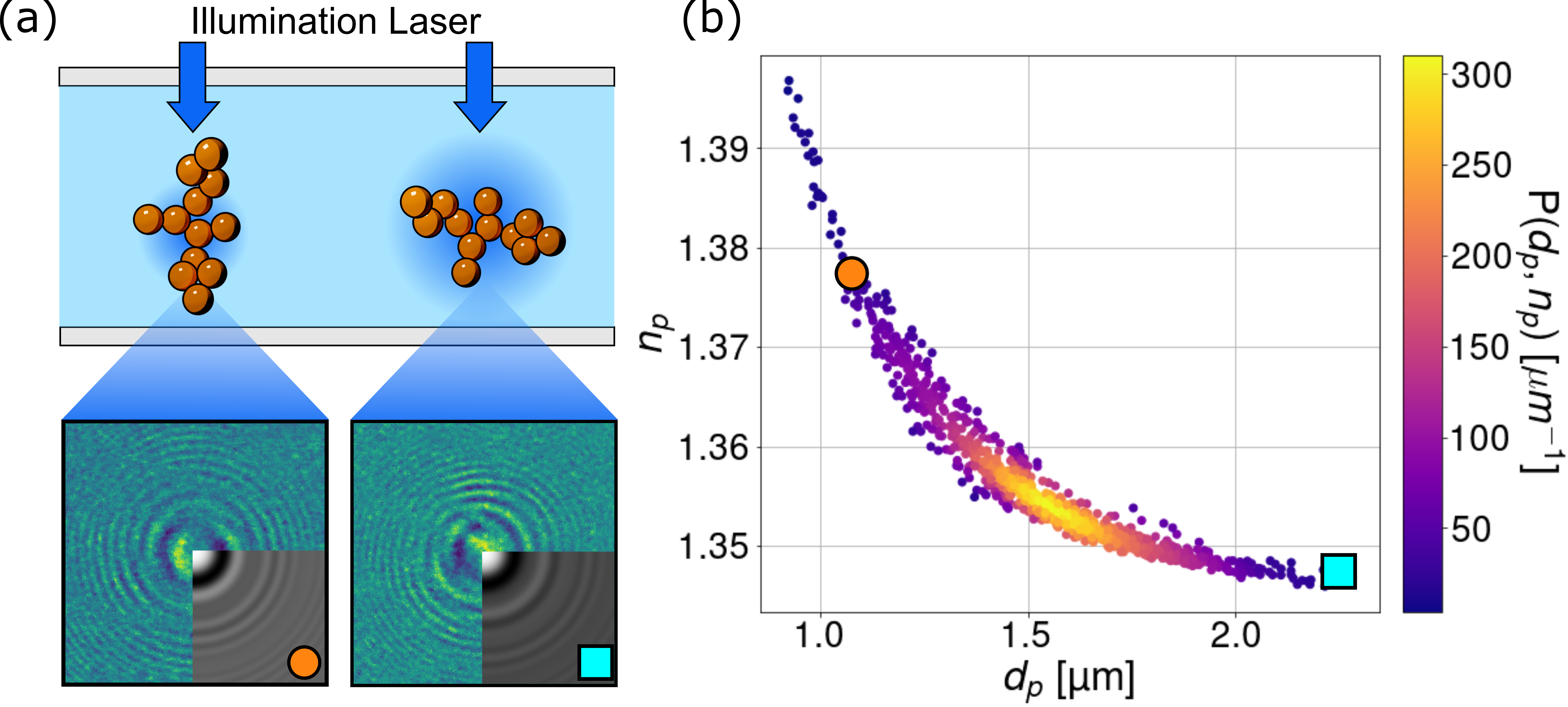}
  \caption{(a) Schematic representation of holographic
    characterization of a diffusing fractal aggregate in a microfluidic
    channel. The aggregate, passing through the illuminating laser beam,
    scatters light that interferes with the rest of the laser beam to
    create a hologram. The two images are experimental holograms of a
    single aggregate in different orientations with fits to Lorenz-Mie
    theory superimposed in the lower-right quadrants.  (b) Effective
    values for $d_p$ and $n_p$ for the aggregate in multiple orientations.
    Each plot symbol corresponds to a single observation of the freely
    diffusing cluster and is colored by the density of observations,
    $P(d_p, n_p)$. Orange and teal symbols denote results from the holograms
    in (a).}
  \label{fig:overview}
\end{figure*}

\section{Holographic particle characterization
  in the effective-sphere approximation}
\label{sec:theory}

In-line holographic microscopy uses a collimated laser beam to
illuminate the sample \cite{sheng2006digital}.  Light scattered by
colloidal particles in the sample interferes with the rest of the
illumination.  An optical microscope magnifies the interference
pattern and relays it to a video camera, which records its intensity.
Each image in the video stream is a hologram of the particles in the
observation volume that encodes information about the particles'
positions and compositions.  That information can be extracted by
fitting a recorded hologram to a generative model for the image
formation process.

We model the illumination as a monochromatic plane wave that is
linearly polarized along $\hat{x}$ and propagates along $\hat{z}$:
\begin{equation}
    \vec{E}_0(\vec{r})
    =
    u_0 \, e^{i k z} \, \hat{x}.
\end{equation}
A small spherical particle located at $\vec{r}_p$ scatters a wave,
\begin{equation}
    \vec{E}_s(\vec{r}) =
    E_0(\vec{r}_p) \,
    \vec{f}_s(k(\vec{r} - \vec{r}_p)),
\end{equation}
that is proportional to the incident field at its position.  The
structure of the scattered wave is described by the Lorenz-Mie
scattering function, $\vec{f}_s(k\vec{r})$
\cite{bohren2008absorption,mishchenko2002scattering,gouesbet2011generalized},
which is parameterized by the sphere's diameter, $d_p$, and its
refractive index, $n_p$.  The superposition of the incident and
scattered waves creates an interference pattern whose intensity is
recorded by the camera \cite{lee2007characterizing}:
\begin{equation}
    b(\vec{r})
    =
    \abs{\hat{x} + e^{-i k z_p} \, \vec{f}_s(k(\vec{r} - \vec{r}_p))}^2,
    \label{eq:model}
\end{equation}
where we have normalized the recorded image
\cite{lee2007characterizing} so that $u_0^2 = 1$.  Distances are
measured relative to the vacuum wavelength of light, $\lambda$,
through the wave number, $k = 2 \pi n_m / \lambda$, in a medium of
refractive index $n_m$. In typical implementations, $\lambda$, $n_m$,
and the magnification of the microscope are treated as fixed
instrumental parameters.  This generative model for the
image-formation process can be fit to a measured hologram by
optimizing $\vec{r}_p$, $d_p$ and $n_p$.  When applied to homogeneous
colloidal spheres, this procedure finds the center of particle with a
measurement error of $\sigma_x = \sigma_y = \SI{2}{\nm}$ and $\sigma_z
= \SI{5}{\nm}$.  The same measurement yields the sphere's diameter
with a precision of $\sigma_d = \SI{2}{\nm}$ and its refractive index
with a precision of $\sigma_n = \num{0.001}$
\cite{krishnatreya2014measuring}.

Equation~\eqref{eq:model} can be generalized to accommodate aspherical
and inhomogeneous particles by selecting a suitable form for the
scattering function \cite{wang2014using}.  Such generalizations,
however, dramatically increase the time and resources needed to seek
optimal solutions to the inverse problem
\cite{barkley2019holographic}.  The effective-sphere approximation
avoids this computational cost by analyzing holograms of aspherical
particles using the Lorenz-Mie scattering function for homogeneous
spheres.  In this case, optimal values for the diameter and refractive
index reflect the properties of an effective sphere encompassing the
particle whose index may be interpreted with Maxwell Garnett
effective-medium theory \cite{markel2016introduction} to obtain
information about the particle's true underlying properties
\cite{cheong2011holographic,hannel2015holographic,wang2016fractal,odete2020role}.
A porous sphere, for example, consists of a base material of
refractive index $n_0$ that comprises a fraction, $\phi$, of its
volume.  The rest of the volume is filled with the fluid medium at
refractive index $n_m$.  The effective refractive index for such a
two-component system is \cite{cheong2011holographic,odete2020role}
\begin{subequations}
\label{eq:np}
\begin{equation}
    n_p = \sqrt{\frac{1 + 2 \phi \, L(n_0/n_m)}{1 - \phi \, L(n_0/n_m)}},
\end{equation}
where the Lorentz-Lorenz function is
\begin{equation}
    L(m) = \frac{m^2 - 1}{m^2 + 2}.
\end{equation}
\end{subequations}

Effective-sphere analysis can be applied to aggregates of
nanoparticles, which tend to be aspherical as well as porous, provided
that the aggregates' asperities are small enough for effective-medium
theory to apply
\cite{wang2016fractal,wang2016protein,fung2019computational}.  The
number, $N = (d_p/d_0)^D$, of monomers of diameter $d_0$ in a cluster
of diameter $d_p$ depends on the aggregation mechanism and is
characterized by the fractal dimension, $D$.  The monomers in such a
cluster occupy a fraction
\begin{equation}
    \phi(d_p) = \frac{N d_0^3}{d_p^3} = \left(\frac{d_p}{d_0}\right)^{D-3}
\end{equation}
of the aggregate's volume, the rest being filled with the medium.
This suggests that the effective refractive index of a fractal
aggregate should scale with its diameter as \cite{wang2016fractal}
\begin{equation}
    \ln\left(
    \frac{L(m_p)}{L(m_0)}
    \right)
    = (D - 3) \, 
    \ln \left(\frac{d_p}{d_0}\right),
    \label{eq:scalingrelation}
\end{equation}
where $m_p = n_p/n_m$ and $m_0 = n_0/n_m$.
Equation~\eqref{eq:scalingrelation} has been used successfully to
assess the fractal dimension of populations of protein aggregates
\cite{wang2016protein}, and nanoparticle clusters
\cite{wang2016fractal}, and has been validated through simulations of
light scattering by fractals \cite{fung2019computational}.  These
numerical studies also suggest that Eq.~\eqref{eq:scalingrelation}
should hold for individual fractal clusters.

\section{Multi-angle holography of fractal clusters}
\label{sec:holographictomography}

As illustrated in Fig.~\ref{fig:overview}(a), a fractal aggregate's
hologram changes as the the aggregate rotates, leading
effective-sphere analysis to settle on different values for the
effective diameter and refractive index.  The data in
Fig.~\ref{fig:overview}(b) were obtained for a single fractal silica
aggregate with each point representing one measurement of the
freely-diffusing cluster.  The two representative holograms in
Fig.~\ref{fig:overview}(a) are called out with large plot symbols in
Fig.~\ref{fig:overview}(b) and generally span the range of effective
properties exhibited by this aggregate in different orientations.
Such a distribution of values can be interpreted with
Eq.~\eqref{eq:scalingrelation} to obtain a tomographic estimate for
the cluster's fractal dimension, $D$.

The tomographic value of $D$ for an individual aggregate can be
compared with the average scaling behavior obtained from
single-hologram analysis of multiple clusters.  This comparison
constitutes a test of the assumption underlying previous holographic
characterization studies \cite{wang2016fractal,wang2016protein} that
all of the fractal aggregates in a population scale in a similar way
and that their scaling behavior can be captured by effective-sphere
analysis.  This test can be made more rigorous by performing
measurements over a range of growth conditions designed to yield
aggregates with different fractal dimensions.

Our model system consists of dispersions of silica nanospheres (Ludox
TMA colloidal silica, Aldrich catalog no.~420859) that are induced to
aggregate through the addition of salt.  The spheres in this system
have mean diameter $d_0 = \SI{20}{\nm}$ as determined by dynamic light
scattering (LS Spectrometer, LS Instruments; Zetasizer Nano ZS,
Malvern Instruments).  The stock dispersion has \SI{34}{wt\percent}
solids, which corresponds to a number density of $c =
\SI{5e16}{\per\milli\liter}$ assuming the density of silica to be
$\rho_{\ce{SiO2}} = \SI{2.2}{\gram\per\cubic\centi\meter}$.  This
stock solution is diluted with deionized water (MilliQ Ultrapure Water
System, MilliporeSigma) and is destabilized by adding \SI{0.1}{M}
\ce{MgCl2} (Sigma-Aldrich, CAS no.~7791-18-6).  Aggregation is allowed
to proceed for \SI{1}{\hour} before the dispersion is further diluted
with deionized water to arrest growth.  Clusters from the diluted
dispersion are then sampled immediately for analysis.

Holographic characterization of populations of aggregates is performed
with a commercial instrument (xSight, Spheryx, Inc.) that
automatically records and analyzes holograms of thousands of particles
by drawing up to \SI{6}{\micro\liter} of sample through a microfluidic
observation volume in a pressure-driven flow.  Each detected particle
is characterized by a single value for the diameter and refractive
index.  The microfluidic channel used for these measurements (xCell,
Spheryx, Inc.) has a \SI{50}{\um} minimum dimension, which means that
the concentration of micrometer-scale particles can be measured
accurately at concentrations up to \SI{e7}{\per\milli\liter} before
neighboring particles' holograms overlap significantly.


Individual particles from each sample also were analyzed in a
custom-built holographic microscope operating at a vacuum wavelength
of $\lambda = \SI{0.447}{\um}$ (Coherent Cube) and an effective system
magnification of $\SI{48}{\nm\per pixel}$.  Samples were contained in
microfluidic chambers formed by sealing the edges of a \#1.5 glass
cover slip to the surface of a standard glass microscope slide with
UV-cured optical adhesive (NOA81, Norland Products).  The \SI{30}{\um}
nominal thickness of these chambers is large enough for an aggregate
to diffuse freely in three dimensions.  The holographic microscope
therefore observes the aggregate in all orientations over time.  The
particle's tendency to diffuse out of the field of view was controlled
using a holographic optical trapping system integrated into the
instrument \cite{obrien2019above}.  This instrument's camera (Flea3
monochrome USB3.1, Teledyne FLIR) records holograms at
\SI{30}{frame\per\second} with an exposure time of \SI{10}{\us}, which
is fast enough to avoid motion blurring
\cite{cheong2009flow,dixon2011holographic}
given the translational and rotational diffusion coefficients
for a micrometer-diameter object,
\begin{equation}
    D_r = \frac{k_B T}{3 \pi \eta d_p} \approx \SI{0.4}{\square\um\per\second} \quad \text{and} \quad
    D_\theta = \frac{k_B T}{\pi \eta d_p^3} \approx \SI{1.3}{\per\second},
\end{equation}
respectively, in a medium with viscosity  
$\eta = \SI{1}{\milli\pascal\second}$.
Single-particle holograms were
analyzed with the open-source \texttt{pylorenzmie} package.

Aggregation at vanishingly low monomer concentration allows clusters
to grow through diffusion-limited aggregation (DLA)
\cite{witten1983diffusion} and yields clusters with fractal dimensions
as large as $D = \num{2.3}$.  Destabilizing a more concentrated
dispersion allows for diffusion-limited cluster aggregation (DLCA),
which leads to more open and spindly structures with fractal
dimensions around $D = \num{1.3}$ \cite{meakin1984topological}.

\begin{figure*}[ht]
\centering
  \includegraphics[width=0.75\textwidth]{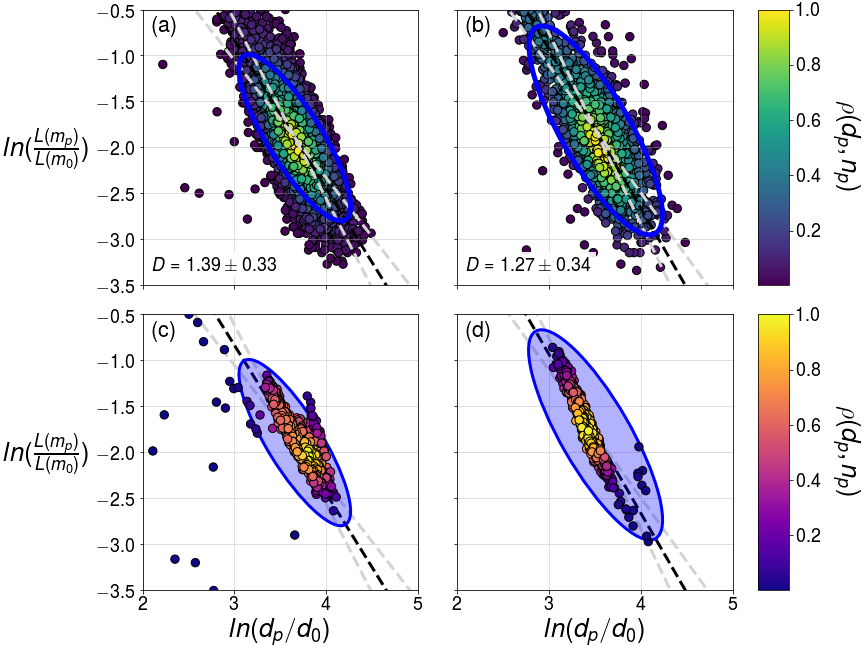}
  \caption{Characterization data for two populations of particles formed
  by aggregating silica nanoparticles from an initial concentration
  of (a) \SI{1.4e15}{nanoparticles\per\milli\liter} and
  (b) \SI{5.6e15}{nanoparticles\per\milli\liter}.
  Each point represents the diameter, $d_p$, and refractive index,
  $n_p$, of a single aggregate, rescaled according to Eq.~\eqref{eq:scalingrelation}. Points are colored by the
  relative density of measurements, $\rho(d_p, n_p)$.
  Best-fit (dashed) lines show estimates for the population-averaged
  fractal dimension together with uncertainty bounds.
  (c) and (d) show multi-angle characterization results for
  individual aggregates sampled from the dispersions in 
  (a) and (b), respectively. The 2-$\sigma$ confidence 
  ellipses and best-fit lines
  from (a) and (b) are reproduced to aid comparison between
  population-averaged and single-cluster results.}
  \label{fig:characterization}
\end{figure*}

The holographic characterization data in Fig.~\ref{fig:characterization}(a) and (b) were obtained
for populations of \num{5000}
silica aggregates grown at starting monomer concentrations of $c=
\SI{1.4e15}{\per\milli\liter}$ and \SI{5.6e15}{\per\milli\liter},
respectively.
Each point represents the measured diameter and refractive index
of a single particle, recast into the form of Eq.~\eqref{eq:scalingrelation} and is colored by
the relative density of measurements, $\rho(d_p, n_p)$.
Both sets of results generally follow the linear
trend predicted by Eq.~\eqref{eq:scalingrelation}.
Aggregates grown at lower monomer concentration have an
inferred fractal dimension of $D = \num{1.39(33)}$ while those grown
at higher concentration have a slightly lower value, $D =
\num{1.27(34)}$, which is consistent with expectations for
three-dimensional DLCA \cite{meakin1984topological}.
These ranges are overlaid as dashed lines in
Fig.~\ref{fig:characterization}(a) and Fig.~\ref{fig:characterization}(b).

Multi-angle characterization of individual aggregates
yields results that are consistent with population-averaged
measurements.
The data in Fig.~\ref{fig:characterization}(c) and (d)
were obtained for aggregates drawn from the same populations
as (a) and (b), respectively.
Points in these scatter plots represent effective diameters and
refractive indexes that are measured over time
as the aggregate undergoes rotational diffusion.
The associated population distributions are reproduced
as 2-$\sigma$ confidence ellipses to facilitate comparison.
In both cases, the fractal dimension of the
single-particle distribution agrees with that of the corresponding
population.

\begin{figure}[ht]
  \centering
  \includegraphics[width=0.9\columnwidth]{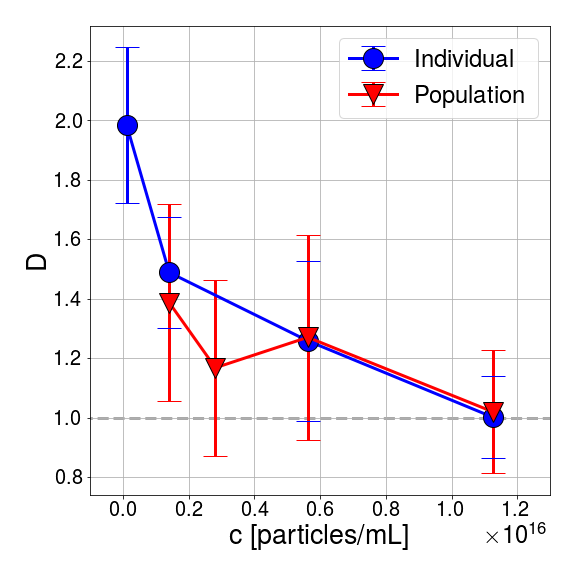}
  \caption{Measured population-average fractal dimension as a function
  of initial nanoparticle concentration.
  Trend lines for both individual aggregates (blue circles)
  and populations of aggregates (red triangles) suggest an
  inverse relationship between concentration of nanoparticles $c$
  and fractal dimension $D$.
  The dashed line at $D=1$ represents the lower physical limit.
  }
  \label{fig:fractaldimension}
\end{figure}

Figure~\ref{fig:fractaldimension} shows how the population-averaged
fractal dimension varies with monomer concentration over two decades
from $c = \SI{e14}{\per\milli\liter}$ to \SI{e16}{\per\milli\liter}.
As expected, the clusters' fractal dimension is anticorrelated with
the initial monomer concentration.
The measured concentration of aggregates is
\SI{e6}{aggregates\per\milli\liter} in the most dilute samples, 
with a typical aggregate containing an estimated \num{5000} monomers. 
This suggests that no more than \SI{5e9}{monomers\per\milli\liter} have
aggregated into clusters that are large enough to be detected.  Most
of the nanospheres therefore either remain in monomeric form or have
formed aggregates smaller than the measurement technique's detection
limit of $d_p \lesssim \SI{500}{\nm}$.
Based on these estimates, the initial
concentration of monomers is high enough to support cluster-cluster
aggregation over the full range of growth conditions considered in
this study and helps to further validate the observed
range of fractal dimensions. 
While the population-averaged data show high variance, limiting the conclusions that can be confidently drawn, the data from the individual aggregate analysis are in agreement with our anticorrelation hypothesis.

\section{Tracking rotational diffusion of a fractal aggregate}

\begin{figure*}[ht]
    \centering
    \includegraphics[width=\textwidth]{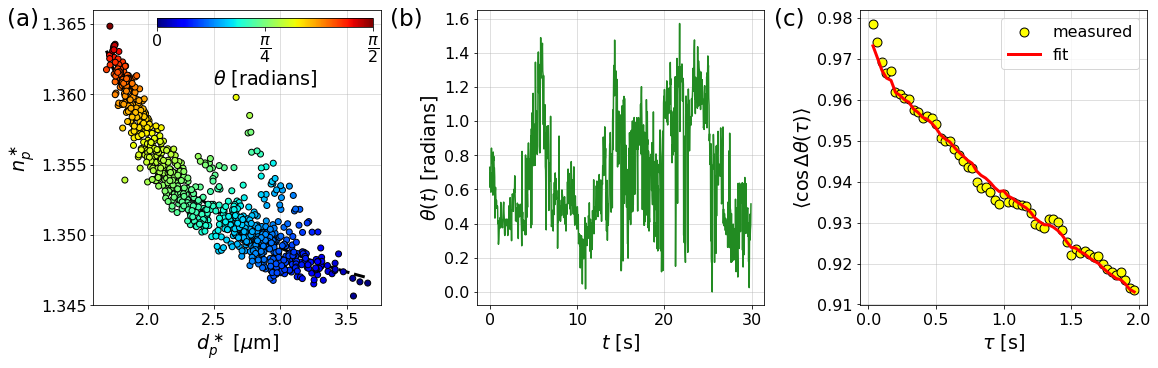}
    \caption{(a) Effective-sphere characterization data
    for the particle from Fig.~\ref{fig:characterization}(c) colored according
    to inferred polar angle, $\theta$.
    (b) Angular trajectory, $\theta(t)$.
    (c) Mean-square displacement of the orientation
    unit vector as a function of lag time, $\tau$,
    together with a fit to Eq.~\eqref{eq:rotationaldiffusion} for the
    particle's rotational diffusion coefficient,
    $D_\theta$, and the uncertainty, $\epsilon$,
    in the measured orientation angle.
    The dashed curve shows the naive fit
    that does not account for measurement errors.}
    \label{fig:angular_diffusion}
\end{figure*}

Values of the diameter and refractive index inferred
from effective-sphere analysis of an aspherical particle
can be used to estimate the particle's instantaneous
polar orientation angle, $\theta(t)$ \cite{altman2021holographic}.
As an illustrative example, we consider the single-particle
data set from Fig.~\ref{fig:characterization}(c).
The largest effective diameter and smallest effective refractive
index is obtained when the particle's semi-major axis lies
along the imaging plane, $\theta = 0$.
The opposite end of the distribution is obtained when the
semi-major axis is aligned with the optical axis,
$\theta = \pi/2$.
The measured points then can be fit
to the parametric form \cite{altman2021holographic}
\begin{align}
    d_p^\ast(\theta) 
    & = 
    \left[\max(d_p^\ast) - \min(d_p^\ast) \right]\sin{\theta} + \min(d_p^\ast), \\
    n_p^\ast(\theta) 
    & = 
    n_m + \frac{\Delta n}{\kappa \, d_p^\ast (\theta) - 1},
\end{align}
where $\max(d_p^\ast) = \SI{3.66}{\um}$ and 
$\min(d_p^\ast) = \SI{1.69}{\um}$ are the 
largest and smallest values of $d_p^\ast(\theta)$ observed in the
characterization data, respectively, $n_m = \num{1.34}$ is the refractive index of water,
and $\Delta n = \num{0.023(1)}$ and 
$\kappa = \SI{1.18(3)}{\per\um}$ are fitting parameters.
The minimum and maximum values of $d_p^\ast$
may be interpreted as the
particle's approximate minor and major axes, which suggests that the particle has an 
overall aspect ratio of \num{2.2}.
The fit is plotted as a dashed curve
in Fig.~\ref{fig:angular_diffusion}(a).
Each data point is assigned the polar angle, $\theta(t)$, that minimizes the
distance from
$(d_p^\ast(t), n_p^\ast(t))$ to the fit curve.
The data points in Fig.~\ref{fig:angular_diffusion}(a)
are colored accordingly.

Following this procedure, 
the sequence of measured effective-sphere
characterization results can be
recast into the angular trajectory
that is plotted in
Fig.~\ref{fig:angular_diffusion}(b).
This trajectory, in turn, can be used to estimate the particle's
rotational diffusion coefficient through the
Stokes-Einstein relation \cite{jain2017diffusing}
taking into account
the measurement error, $\epsilon$, in the
particle's orientation, $\theta(t)$:
\begin{equation}
\label{eq:rotationaldiffusion}
    \avg{\cos\Delta\theta(\tau)}
    =
    \exp(-D_\theta \tau)
    + \epsilon \, \sqrt{2 \avg{\sin^2\Delta\theta(\tau)}}.
\end{equation}
Here, $\Delta\theta(\tau) = \theta(t + \tau) - \theta(t)$
and angle brackets indicate averages over $t$.
Values of
$\avg{\cos\Delta\theta(\tau)}$ computed
from $\theta(t)$ are plotted as discrete
points in Fig.~\ref{fig:angular_diffusion}(c).
Fitting these points to Eq.~\eqref{eq:rotationaldiffusion} yields
the solid curve, 
$D_\theta = \SI{0.019(1)}{\per\second}$.
The corresponding particle diameter, $d_p =
\SI{4.1(1)}{\um}$,
is consistent the optically-inferred 
major axis of the particle, which is
expected for rotational diffusion of fractal clusters \cite{lattuada2004rotational}.
The fit also yields an uncertainty 
of $\epsilon = \SI{5.3(1)}{\degree}$ in
the effective-sphere measurement of this aggregate's orientation angle.

The effective-sphere approach to
holographic rotation tracking works
best for clusters with the largest aspect ratios, typically
those with the smallest fractal dimensions.
Reasonable outcomes for particles such as the
typical example in Fig.~\ref{fig:angular_diffusion}
serve to validate the overall structural analysis
of this common class of colloidal particles.
It also suggests that such particles can be
recognized and differentiated from 
other sorts of particles, such as colloidal spheres, without
invoking \emph{a priori} knowledge of their
structure and composition.

\section{Toward holographic tomography}

The methods we have described use holograms
recorded over a range of orientations to 
describe the overall structure of fractal
colloidal aggregates under the assumption that
they actually are fractal.
This approach can be extended to obtain more
detailed information about the inner structure
of such objects.
The volume of the bounding effective sphere can
be subdivided into voxels characterized by 
a set of refractive indexes, $\{n_j\}$, that can
be used to compute the scattering T matrix for
the particle \cite{mishchenko2002scattering}.
The T matrix then can be used to compute the
particle's hologram as a function of
orientation, $\vec{\Omega}$.
In principle, $\{n_j\}$ can be selected to optimize
agreement between these synthetic holograms 
and the set of experimental holograms in random
orientations. The optimized model not only would
provide direct insights into the distribution of
matter within the particle, but also could
be used to estimate $\vec{\Omega}$ for the cluster.
This optimization scheme currently is too computationally
expensive for practical implementation.
It may be possible to accelerate convergence 
by adapting
machine-learning techniques for analyzing
sequences of single-particle holograms
\cite{midtvedt2021fast}.

\section{Discussion}
\label{sec:discussion}

Our results demonstrate that effective-sphere analysis of holographic
microscopy data can provide useful information about the morphology of
individual fractal clusters.
Rather than relying on population
averages, this approach draws insights from the distribution of
effective-sphere characterization results obtained as the particle
rotates in three dimensions.
Rather than yielding particular values
for the diameter and refractive index,
effective-sphere analysis of a rotating
fractal aggregate yields estimates
for the particle's fractal dimension,
its aspect ratio, 
and its rotational diffusion coefficient.
When applied to aggregates of well-characterized monomers
this information
offers insights into the mechanisms for those clusters' formation.

The wealth of information provided by
holographic particle characterization
should be useful in
any industry where particle properties affect
commercial success.
Holographic characterization probes particles in their
native media without extensive sample preparation,
which is an advantage relative to standard
techniques such as electron microscopy.
The ability to differentiate fractal aggregates from
other types of particles in heterogeneous real-world
dispersions sets holographic characterization apart
from conventional light-scattering techniques.
Holographic characterization is fast
enough, moreover, for applications in process control and
quality assurance.
The population-averaged measurements in Fig.~\ref{fig:fractaldimension}(a) and (b) 
required less than \SI{15}{\minute} each, and
multi-angle characterization of a colloidal
cluster takes just a few seconds thanks to
the recent introduction of software for real-time 
hologram analysis \cite{altman2020catch}.

\section*{Funding}
This work was primarily supported by the SBIR program of the National
Institutes of Health under Award Number R44TR001590.  The Spheryx
xSight, LS Instruments LS Spectrometer, and Malvern Zetasizer Nano ZS
were acquired as shared instrumentation by the NYU MRSEC with support
from the National Science Foundation under Award Number
DMR-1420073. The custom holographic characterization instrument was
constructed with support from the MRI program of the NSF under Award
Number DMR-0923251.

\section*{Acknowledgments}
We are gradeful to Prof.\ Andrew Hollingsworth for assistance with the
preparation and light-scattering characterization of the fractal
aggregates used in this study.

\section*{Disclosures}
DGG is a founder of Spheryx, Inc., the company that manufactures the
xSight particle-characterization instrument used in this study.

\section*{Data availability}
Data underlying the results presented in this paper are not publicly
available at this time but may be obtained from the authors upon
reasonable request.


\end{document}